\documentclass[11pt]{article}
\usepackage{fullpage}
\usepackage{subcaption}
\usepackage{comment}

\usepackage{cite}
\usepackage{amsmath,amssymb,amsfonts}
\usepackage{algorithmic}
\usepackage{graphicx}
\usepackage{textcomp}
\usepackage{xcolor}
\usepackage{hyperref}
\usepackage{soul}
\usepackage{flushend}

\newcommand{\note}[1]{{\color{red}\textit{#1}}}

\newcommand{\TOOL}{DHP}
\begin{document}

\title{NavP:  Enabling Navigational Programming for Science Data Processing via Application-Initiated Checkpointing

}

% Commenting out for the double blind submission

% \author{
% \IEEEauthorblockN{Lei Pan}
% \IEEEauthorblockA{\textit{Jet Propulsion Laboratory, California Institute of Technology}} \\
% Pasadena, USA\\
% lei.pan@jpl.nasa.gov\\
% \and
% \IEEEauthorblockN{Twinkle Jain}
% \IEEEauthorblockA{\textit{Northeastern University}} \\
% Boston, USA\\
% jain.t@northeastern.edu
% }
% \author{author(s) omitted}

\author{
Lei Pan \\
Jet Propulsion Laboratory \\
California Institute of Technology \\
Pasadena, USA\\
lei.pan@jpl.nasa.gov\\
\and
Twinkle Jain\thanks{\noindent This work was partially supported
by National Science Foundation Grant OAC-1740218 and a grant from
Intel Corporation.} \\
Northeastern University \\
Boston, USA \\
jain.t@northeastern.edu
}
\date{}

\maketitle

\begin{abstract}
Science Data Systems (SDS) handle science data from acquisition through processing to distribution. They are deployed in the Cloud today, and the efficiency of Cloud instance utilization is critical to success. Conventional SDS are unable to take advantage of a cost-effective Amazon EC2 spot market, especially for long-running tasks. Some of the difficulties found in current practice at NASA/JPL are: a lack of mechanism for app programmers to save valuable partial results for future processing continuation, the heavy weight from using container-based (Singularity) sandboxes with more than 200,000 OS-level files; and the gap between scientists developing algorithms/programs on a laptop and the SDS experts deploying software in Cloud computing or supercomputing. %%; and the use of complex HDF/newCDF readers to access remote data.

All of the above difficulties can be ameliorated if the process carrying out computations on the scientist's laptop can be directly migrated among multiple nodes in the Cloud or on the supercomputer. The key idea is to allow the scientist programmers to directly navigate the computations to the data, or help them move the remaining computations to a new Cloud instance when the old one is taken away.
We present a first proof-of-principle of this using NavP (Navigational Programming) and fault-tolerant computing (FTC) in SDS, by employing program state migration facilitated by Checkpoint-Restart (C/R). NavP provides a new navigational view of computations in a distributed world for the application programmers. The tool of DHP (DMTCP Hop and Publish) we developed enables the application programmers to navigate the computation among instances or nodes by inserting $\mathtt{hop(destination)}$ statements in their app code, and choose when to publish partial results at stages of their algorithms that they think worthwhile for future continuation. The result of using DHP is that a parallel distributed SDS becomes easier to program and deploy, and this enables more efficient leveraging of the Amazon EC2 Spot market. This technical report describes a high-level design and an initial implementation.

NASA/Jet Propulsion Laboratory acquires massive amounts of observatory instrument data in their missions from planetary (e.g., Mars Exploration, Cassini, Rosetta, Psyche) to earth science (e.g., Jason, OCO, NISAR, SMAP, MAIA). This instrument data is then processed into science data on massively parallel distributed computing facilities such as the Cloud and supercomputers. Innovations in SDS programming are the key to the success of JPL missions.

%%% \TODO{Highlight NASA/JPL involvement for importance and novelty \\
%%%  Also, highlight \TOOL{} in the abstract}
% What - we are enabling NavP and FT via C/R
% Why - SDS are unable to leverage EC2 spot 
% How - By combining NBS and C/R
\end{abstract}

% \begin{IEEEkeywords}
% checkpoint-restart, DMTCP, NavP, Fault-Tolerant Computing, Science Data Systems
% \end{IEEEkeywords}
\vspace{-2.2mm}
\section{Introduction}
Large-scale Science Data Systems (SDS) like Hybrid Cloud Science Data System (HySDS)~\cite{hua:2021} are widely used in various critical NASA missions~\cite{hua:2021}. Cloud computing is elastic and scalable, making it an ideal choice for SDS. Another platform for SDS is cluster computers, an example of which is NASA's petascale supercomputer, Pleiades, which has seen heavy investment over the last decade.

Our first goal is to help the scientist programmers to go directly from algorithms on paper to final production runs in SDS, without much help from an outside specialist in SDS.
By allowing the scientist-programmers to directly program and immediately test and deploy (without much help of the SDS expert), the edit-test-debug-production cycle of software development will be greatly accelerated.
Another goal is to facilitate high performance and effective resource leveraging.

However, in the currently available SDS, there are three major problems that prevent us from achieving this goal.
\begin{enumerate}
  \item 
Long-running tasks are not readily broken down into smaller ones to leverage the Amazon EC2 spot market. This makes it hard or impossible to exploit the steep discounts of the EC2 spot market.
It is the temporal aspect of atomic jobs.
  \item
  The prevailing container-based approach to deploying programs and moving computation to data is heavyweight, in that it moves much more loaded data than the strictly necessary program state. In the case of a Singularity sandbox, which is the virtualization technique adopted on Pleiades, the run-time environment being moved includes 200,000+ OS-level files. Furthermore, container images, in which app programs are installed, take a long time to build, and require knowledge and skills in SDS deployment, such as Terraform scripting, AWS instance management, Docker/Singularity build scripting, and Jenkins pipeline scripting, which scientist programmers in general are not familiar with.  One would always have to work with an SDS expert in virtualizing and deploying the application-level programs, and that is even true for doing version updates. Part of the problem is mixing different levels of abstraction for different concerns, such as app algorithms vs\hbox{.} details of data distribution. Therefore, unnecessary burdens are placed on the scientist-programmer's shoulders. This is the spatial aspect of atomic jobs.
  \item Only embarrassingly parallel algorithms (e.g., MapReduce) are easily programmable. In general, parallel programming is a non-trivial task, especially in a distributed environment. This is due to the fact that prevailing programming methodologies provide a stationary view of the distributed systems where stationary processes work with each other using messages. Using an analogy in traveling as an example, there are two views, namely arrivals-and-departures in the train stations/airports and an itinerary in a traveler's hand. These two views both capture the exact same information, spatial and temporal, about a traveler. Our proposal is that, just as a traveler should be using an itinerary to travel with, distributed parallel programming should be using the navigational view. The reality though is that all prevailing programming paradigms today are as awkward as traveling using the arrivals-and-departures view. Our proposal comes with a grand challenge: how to make NavP almost as efficient as message passing, for all major programming languages. This would mean optimizations at the underlying DMTCP level: (1)~To travel to a remote node carrying only the data needed for future computation; and (2)~To avoid moving code, including both app level code and run-time environment level code, e.g., OS code, python modules, shared libraries, to visit a node more than once.
\end{enumerate}
% \TODO{NASA JPL missions such as ARIA, SMAP, SWOT, NISAR, OCO-2 are among the users among of the HySDS system.}

We leverage checkpoint-restart (C/R) to enable Navigational Programming (NavP)~\cite{pan2004navp} in order to facilitate high performance, effective resource leveraging, and ease of use for scientist programmers.

Navigational Programming (NavP) was originally developed in~2004~\cite{pan2004navp}.  The advantage of NavP for distributed parallel programming is that it brings the computation to the data using an intuitive navigational view of the distributed programming environment, as opposed to the conventional view of message passing, which in general would force massive restructuring of the original sequential algorithms. NavP has not yet been widely adopted because process migration for multiple programming languages across the board is difficult.  Furthermore, making process migration as efficient as message passing is a grand challenge at the underlying system level.  However, we argue that the time has now come for this novel paradigm.  This work presents an architecture that leverages a principled implementation relying on checkpointing in order to enable NavP.

%%% We discuss these concepts in detail in the next section.

The next five items each present background on the practical requirements of Scientific Data Systems (SDS), followed by a question that motivates the goals of the NavP project.  This paper answers those questions in the positive, by showing a first proof-of-principle for NavP.

%%% \note{Introduction should include a number of motivations mentioned by Lei}
%%% \TODO{@Lei: correct and add more information in the paragraphs below}

\begin{enumerate}
\item Large-scale software development typically uses Continuous Integration (CI) and Continuous Deployment (CD) to check in, build, test, and deploy incremental changes in code development. In practice, new versions of the code are built in and deployed using Docker/Singularity images. The overhead of this approach is the virtual run-time environment, i.e., some 200,000 files from the operating system, and cannot be ignored since CI/CD happens continuously and frequently as scientists update their algorithms. This overhead can become overwhelming in some cases. For example, NASA's supercomputer Pleiades uses Singularity instead of Docker for security considerations.  A Singularity sandbox with more than 200,000 small files will take an enormous amount of time to handle (untarring or deleting) on Pleiades' Lustre file system~\cite{smallfiles}.

Can programs be made to migrate themselves? If so, the virtual containers are then deployed only once, and the subsequent CI/CD operations deploy only the software applications.

\item Scientific application developers are usually scientists who write code on their laptops or desktops, where data is a small sample, and where processing power is limited. Going to the Cloud or supercomputer involves CI/CD administered by SDS experts. So if a scientist develops a new algorithm that works on laptops with small input data, it will require days or even weeks for the scientist to try it out on a larger scale. This is not acceptable.

Can we convert the Cloud or supercomputer into a virtual environment that acts as an extension of the scientist's laptop?  If so, the scientist can test their new algorithms at any time, even at midnight when a new idea suddenly comes to them. The program's self-migration envisioned here enables the scientists to do just this.

\item Webification tools such as JPL's Pomegranate~\cite{w10n_2018} use the http protocol and web browsers to open HDF/netCDF files remotely and transfer only what is needed by the clients. This avoids downloading entire files in cases where only a fraction is useful. Over the years, people have developed HDF/netCDF readers that work on local files efficiently.

Can we can migrate the readers to where the HDF/netCDF files are? If so, file access can be made just as efficient as using HDF/netCDF readers. And at the same time, all of the existing code can be reused. If this can be done, there would be no need to develop or learn to use new tools such as Pomegranate.

%%% We can hop there to have local access
%%% \note{Notes from the meeting:} If an application wants to interact with file(s) over network then there are two ways: 1) open files remotely and access the data (for example Zhangfan's tool pomegranate use http to open remote HDF/netCDF (file format for satellite's data) files) 2) transfer files from remote to local and access them locally. With the help of C/R, this scenario can be changed. The main idea is to migrate the computation where the locus of large data is present. It is common for SDS to interact with a small number of files at a time from the large pool of files. C/R help in improving this scenario two-fold: first, bring computation to near pool of data and access them locally; second, checkpoint along with associated files to move computations with only required files.

%%% Active messages also combine computation and communication. However, they were designed for micro services not navigating full applications.

\item Over the past decades, people spent tremendous effort developing techniques for web services. Corba, JWS, Flask are among the tools provided. These use data in some format, such as xml or json to facilitate communication between the client and the server.

Can programs migrate across the server and client run-time environments?  If so, then the right way to facilitate remote communication will be through the familiar program variables, and programmers will not need to learn anything new.

\item In physics there are two different views of observing moving phenomena:  the Eulerian and the Lagrangian views. These correspond to the arrivals-and-departures (Eulerian) information shown on the screens of airports or train stations, versus the itinerary (Lagrangian) held in the travellers' hand. Computers were originally invented without networking. Hence processes typically are not migrated through the network. So programming using the Eulerian view has been the natural view in programming: all program lines describe local, stationary computation activities, and when a remote computer service is needed, message passing or its variant is used. In earlier work, process self-navigation is proposed as the solution to describe distributed computation --- following its locus of computation in the Lagrangian view~\cite{pan_views_PDCS03}. The result is the Distributed Sequential Computing (DSC) program model from 2001~\cite{pan_DSC_2001}. And multiple DSC's are then synchronized to carry out parallel computation using the idea of a Mobile Pipeline (2005)~\cite{pan_mobile2005}. Scalable performance and easy of programming has been demonstrated in this way, for notoriously hard-to-parallelize numerical algorithms.

In the modern networked world, can NavP now be made a first-class object at the operating system level?  Or can NavP use a development toolkit, such as DMTCP, which in theory can be done as efficiently as message passing, by switching between the two different views? This would be used to describe exactly the same physical phenomenon, using the same amount of message passing, along with some small implementation overhead.

%%% \note{notes from the meeting:} Lei explains that computers were not invented with process migration ability. It it were to support process migration then Eulerian view would not be default. Rather the default approach would be to pass computation as oppose to data in message-passing architecture. The other important factor to consider is that the world is full of heterogeneous systems in terms of both hardware and software. If a single CMI is compatible over even a small subset of heterogeneous systems, it would be a big achievement.

\end{enumerate}

\section{Background}
\label{sec:background}

\subsection{SDS}
%%% \note{Twinkle: I believe it'd be good to understand what is SDS and what are prominent SDS in NASA/JPL. This would clarify that we are talking about real-world applications and how big the audience and the impact of this work would be.}

\begin{figure}[ht!]
  \centering
  \begin{minipage}{0.48\textwidth}
    \centering
    \includegraphics[width=\textwidth]{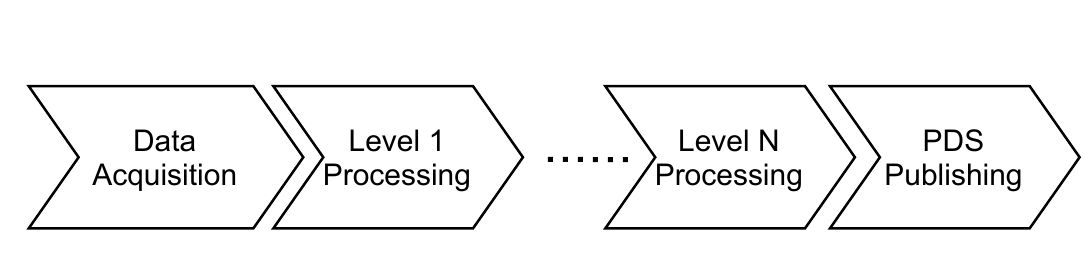}
  \caption{Data processing phases in SDS}
  \label{fig:sds-phases}
\vspace{-1.5em}
  \end{minipage}
\end{figure}

Science Data Systems (SDS) handle science data from acquisition through processing to distribution. This makes a generic SDS a multi-stage workflow, as depicted in Figure~\ref{fig:sds-phases}. Science values are being put into the data as higher level processing are being advanced to. Typically, high level processing requires capabilities involving petascale data storage and processing power. 

% \vspace{-4.5mm}

\subsection{Fault-Tolerant Computing in the Cloud}
\label{subsec:s11}

Amazon provides EC2 spot instances, which offer spare compute capacity in the AWS Cloud, available at steep discounts (90\% savings), but they can be taken away at any time. In the meantime, each atomic task can take hours to finish. Our strategy is to break the original task into smaller pieces using checkpointing and introduce Fault-Tolerant Computing. So, the ``remaining'' computation can be brought to and restarted on a new instance after the old instance disappears.

\subsection{NavP: Navigational Programming}
\label{subsec:s12}

A distributed parallel system is not directly programmable by scientist-programmers. One would always have to work with an SDS expert in virtualizing and deploying the application-level programs, and even when doing version updates. The levels of abstraction for different concerns, application algorithms vs\hbox{.} details of distribution, are coupled. Scientists develop original algorithms that carry out numerical calculations toward data in the form of abstract data structures and variables. The reality of petascale distributed data over the networked systems forces the scientists to work closely with SDS experts to restructure their original algorithms into actual code (e.g., client and server code), install the code in virtual boxes (e.g., Docker images), and deploy the programs in virtual containers onto computer nodes or Cloud instances.
%%% \TODO{@Lei: could you please rewrite the previous line for more clarity?}  
Therefore unnecessary burdens are placed on the application developers\textquotesingle\ shoulders. For applications that are not by nature embarrassingly parallel, this task is difficult, if not impossible. NavP was introduced to address these difficulties~\cite{pan2004navp}. A new view of distributed programming, namely the NavP view, is introduced. In this view, the description of a computation follows its locus to where the large data is found~\cite{pan_views_PDCS03}. This is done by inserting  $\mathtt{hop()}$ statements in the original sequential code. A $\mathtt{hop(dest)}$ statement pauses the computation, collects all the program/thread state, migrates to the $\mathtt{dest}$ node, and resumes computation.

\subsection{Application-initiated transparent C/R}
\label{subsec:s13}

We use the DMTCP (Distributed MultiThreaded CheckPointing)~\cite{ansel2009dmtcp} package for C/R. It transparently checkpoints computations in user-space. It saves a copy of the program state (called a Checkpoint Memory Image (CMI)) to disk and resumes the process later.
It requires no modifications to the user application code nor to the operating system.

The DMTCP plugins~\cite{dmtcp-openproc-2013} and hooks for application-initiated checkpointing provide a flexible way to introduce add-on behaviors around DMTCP events~\cite{ansel2009dmtcp}.

We develop \TOOL \space (DMTCP Hop and Publish), a new Python tool, around DMTCP. This provides two utilities around the checkpoint and restart: a)~\texttt{hop(dest)} --- i.e., responsible for generating CMI on the source node, migrating to, and resuming CMI on the destination node; and b)~\texttt{publish(dest, status)} --- i.e., responsible of publishing job's status and result on the destination node.

The next section describes how to enable $\mathtt{hop()}$ and FTC using \TOOL \space and web services for program state migration.

\section{The NavP Bridging Services (NBS)}
\label{sec:s2}

\subsection{The NavP Bridging Services (NBS) with \TOOL}
\label{subsec:s21}
\begin{figure}[ht!]
  \centering
  \begin{minipage}{0.50\textwidth}
    \centering
    \includegraphics[width=\textwidth]{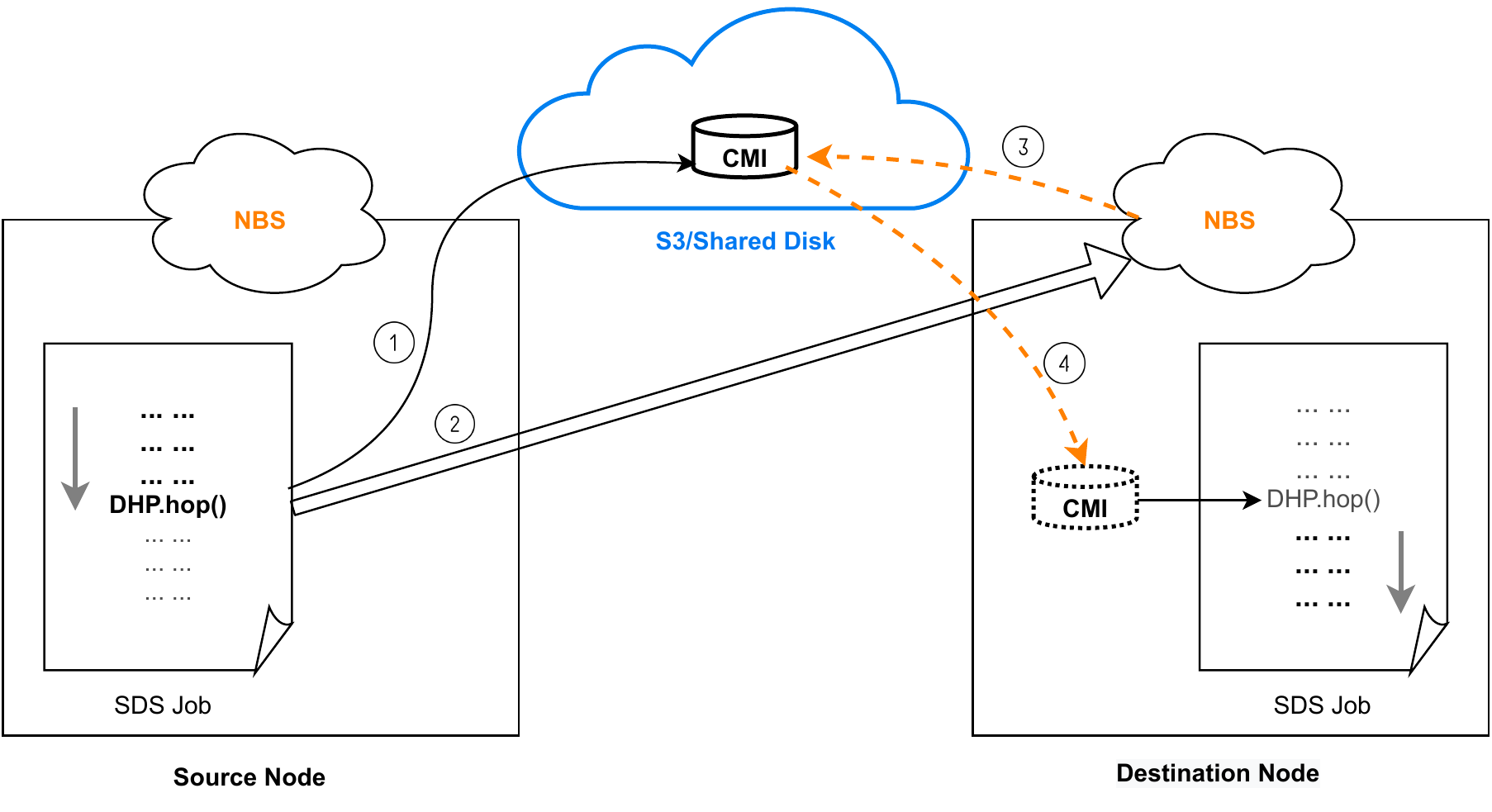}
  \end{minipage}
  \caption{Overview of NavP Bridging Service workflow}
  \label{fig:sds-nbs}
\end{figure}

  The NavP Bridging Services (NBS) run on each compute node and serve the client requests from the application processes. The NBS service for communication and data migration (i.e., $\mathtt{svc/hop}$), running on both source and destination nodes, enables \TOOL \space to migrate CMI(s) between two nodes.
  We provide an overview of the workflow of a NBS in Figure~\ref{fig:sds-nbs}. In the figure, there is a source node \textit{i.e.,} $\mathtt{Host~A}$ and a destination node \textit{i.e.,} $\mathtt{Host~B}$ with NBS services running on each node. Now, an application process running on $\mathtt{Host~A}$ can call the $\mathtt{\TOOL.hop(B)}$ utility, which in turn initiates checkpoint that generates CMI(s) as well as a restart script, and then calls the $\mathtt{svc/hop}$ NBS service running on $\mathtt{Host~B}$ before it terminates itself. The $\mathtt{svc/hop}$ NBS service on $\mathtt{Host~B}$ copies the CMI and restart script from $\mathtt{Host~A}$ to $\mathtt{B}$, and runs the restart script to resume the computation. This way \TOOL~is able to utilize existing data migration utility.
%   This is how process self migration (e.g., $\mathtt{\TOOL.hop()}$) is implemented.
% \vspace{-4.5mm}
  
\subsection{NBS to enable NavP}
\label{subsec:s22}

We use $\mathtt{\TOOL.hop(dest)}$ and $\mathtt{svc/hop}$ to enable NavP.
Pseudocode for the $\mathtt{\TOOL.hop(dest)}$ is in Figure~\ref{code:dmtcp_hop}, in which S3 means some shared disk volume, either 
in an S3 bucket or bound to the containers. Pseudocode for $\mathtt{svc/hop}$ is in Figure~\ref{code:svc_hop}.

\begin{figure}[!ht]
%%% \vspace{0.1in}
\begin{center}
%%% \begin{minipage}{1.5in}
\begin{center}
\mbox{\input{code_dmtcp_hop.tex}}\\[0.3em]
\end{center}
%%% \end{minipage}%
\hspace{\fill}%
\caption{Pseudocode for the $\mathtt{\TOOL.hop(dest)}$ utility}
\label{code:dmtcp_hop}
\end{center}
\vspace{-1.5em}
\end{figure}

\begin{figure}[!ht]
%%% \vspace{0.1in}
\begin{center}
%%% \begin{minipage}{1.5in}
\begin{center}
\mbox{\input{code_hop_service.tex}}\\[0.3em]
\end{center}
%%% \end{minipage}%
\hspace{\fill}%
\caption{Pseudocode for the $\mathtt{svc/hop}$ service}
\label{code:svc_hop}
\end{center}
% \vspace{-1.5em}
\end{figure}

\subsection{NBS to enable FTC in SDS}
\label{subsec:s23}

When jobs in SDS are treated as atomic operations, they can be either ``new'' before the run, or ``finished'' after the run. Any interrupted jobs return to the ``new'' status. A new job has input datasets, and a finished job has products. For the sake of brevity, we ignore more sophisticated situations, such as a ``running'' status, in which a job can have both input datasets and partial products. These additional combinations can be handled in real life applications using the same principle.

The key idea is to introduce a new job status, called ``ckpt'', in which the CMI is treated as a ``special product''. We implement three services to handle the following jobs:
\textbf{(1)}~$\mathtt{svc/list\_jobs}$: this returns all jobs with their $\mathtt{job\_id's}$ along with their statuses, such as those shown in Figure~\ref{code:jobs}.

\begin{figure}[!ht]
%%% \vspace{0.1in}
\begin{center}
%%% \begin{minipage}{1.5in}
\begin{center}
\mbox{\input{code_jobs.tex}}\\[0.3em]
\end{center}
%%% \end{minipage}%
\hspace{\fill}%
\caption{Sample list of jobs with job-ID and status}
\label{code:jobs}
\end{center}
% \vspace{-1.0em}
\end{figure}

\textbf{(2)}~$\mathtt{svc/get\_job}$: this returns the status of a job given its $\mathtt{job\_id}$, or the next job that is not finished when no $\mathtt{job\_id}$ is provided.
\textbf{(3)}~$\mathtt{svc/publish\_job}$: this publishes jobs with two possible statuses: ``ckpt'', in which case the CMI and restart script are uploaded, and ``finished'', in which case the final product is uploaded.

A \TOOL \space utility, $\mathtt{\TOOL.publish(dest, status)}$, is listed in Figure~\ref{code:dmtcp-publish}.  This is similar to $\mathtt{\TOOL.hop(dest)}$. An application uses $\mathtt{\TOOL.publish()}$ to publish with two possible statuses: ``ckpt'' (checkpoints and calls $\mathtt{svc/publish\_job}$ service with a ``ckpt'' status) or ``finished'' (calls $\mathtt{svc/publish\_job}$ with a ``finished'' status). The ``dest'' is the job scheduler service.

\begin{figure}[!ht]
%%% \vspace{0.1in}
\begin{center}
%%% \begin{minipage}{1.5in}
\begin{center}
\mbox{\input{code_dmtcp_publish.tex}}\\[0.3em]
\end{center}
%%% \end{minipage}%
\hspace{\fill}%
\caption{Pseudocode for $\mathtt{\TOOL.publish(dest, status)}$ utility}
\label{code:dmtcp-publish}
\end{center}
% \vspace{-1.5em}
\end{figure}

\section{Experiments: Proof of Concept}

Our experiments are designed to answer the following questions:
\begin{enumerate}
\item How to enable NavP in SDS?
    
The test case application chosen is called ``the co-location of satellite observation data''.  Specifically, the data from the instrument VIIRS (Visible Infrared Imaging Radiometer Suite) is mapped to the geometry of CrIS (Cross-track Infrared Sounder)~\cite{rs8010076,QY2021,fetzer2016}. The first experiment has the pseudocode listed in Figure~\ref{code:app}, in which the calls to $\texttt{svc/publish(``ckpt")}$ are where we checkpoint and where the CMIs are published as ``partial products''.  So when the execution fails for any reason at any time, restart will happen from the most-recent-checkpoint. The application programmer is thus in control of where and how frequently checkpoints happen. The last call to $\texttt{svc/publish(``finished")}$ publishes the real product.
% The $\mathtt{\TOOL.restart()}$ plugin executes the DMTCP restart script with code similar to what is listed in Fig.~\ref{code:svc_hop}.

\begin{figure}[!ht]
%%% \vspace{0.1in}
\begin{center}
%%% \begin{minipage}{1.5in}
\begin{center}
\mbox{\input{code_app.tex}}\\[0.3em]
\end{center}
%%% \end{minipage}%
\hspace{\fill}%
\caption{Pseudocode for JPL application of VIIRS/CrIS co-location.}
\label{code:app}
\end{center}
% \vspace{-1.5em}
\end{figure}

The second experiment assumes that the input data granules are available on a remote server. So the application
needs to first hop to read the input data. Assuming further that this remote server is only for hosting
data (input data and output product), the application hops back to the client server to carry out co-location
matching, before hopping to the remote server again to publish the final product. The pseudocode listed in 
Figure~\ref{code:app_hop} therefore has three $\mathtt{hop()}$ statements inserted in the original code.

\begin{figure}[!ht]
%%% \vspace{0.1in}
\begin{center}
%%% \begin{minipage}{1.5in}
\begin{center}
\mbox{\input{code_app_hop.tex}}\\[0.3em]
\end{center}
%%% \end{minipage}%
\hspace{\fill}%
\caption{Pseudocode for NavP program of the JPL application of VIIRS/CrIS co-location.}
\label{code:app_hop}
\end{center}
\vspace*{-0.25cm}
\end{figure}

\item What is the performance overhead of DMTCP checkpointing and restart? How does this overhead, which is
primarily incurred by disk I/O, from writing out and loading up the CMIs, as compared to the cost of migrating the CMIs 
over the network?

There are two experimental environments, chosen for two different purposes. The first is a single desktop computer running the Linux operating system, with an Intel Xeon 4GHz CPU with 8 cores and 32GB of physical memory. Two virtual NBS containers are run on the same computer, making process migration to happen virtually remotely, but physically only locally. This removes networking cost from the equation, making it ideal for assessing the cost and impact of CMI from checkpointing done by DMTCP. The second system is the Amazon Cloud with m4.4xlarge instances and AWS S3 buckets for storage. This is the real world for JPL SDS, in which the network and S3 performance are all taken into consideration.

\end{enumerate}

We have implementations of all the pseudocode listed in this report working.  From this,
we quickly realized that our general purpose DMTCP needs to be optimized before further
experiments can be meaningfully carried out. This is because as a general purpose C/R
tool, DMTCP currently puts everything our app needs from the run-time environment, e.g.,
python modules and shared libraries, into the CMIs. As a result, the cost of disk I/O
and network transfer of CMIs overshadows the cost of numerical computation
in the app. The Docker/Singularity containers to which the app does hop() or publish(),
are assumed to have identical run-time environments, with the same
python/modules/shared libraries etc. So DMTCP can conveniently skip checkpointing anything related to the run-time environment.  Upon arriving at a remote container, the DMTCP restart shell script will load the modules
and shared libraries from the local installation. This will make the CMIs much more lightweight than they currently are.

Hence, our immediate future work includes a special-purpose, NavP-oriented DMTCP that will be optimized for navigational programming.

\section{Discussion}
In this section, we answer following questions related to the current work.

\bigskip\noindent
\textbf{Question 1.} How to predict when to checkpoint considering spot instance can take their resources back anytime?

Amazon EC2 spot instances are cost-effective resources but can be reclaimed with just two minutes of instance termination notice. This is not sufficient time to initiate, complete, and migrate CMIs for a large-scale job. Jangjaimon et\hbox{.}~al~\cite{jangjaimon2013effective} takes account of past resource reclaim events to predict a checkpoint interval for their adaptive checkpoint scheme. However, the result of an incomplete job is equivalent to no result for an SDS job. Therefore, prediction of the next resource reclaim event would not benefit the user significantly.

\bigskip\noindent
\textbf{Question 2.} What are some advantages of application-initiated checkpoint over conventional periodic checkpoint?

Fixed-interval periodic checkpoints work well where jobs are not considered atomic, and where intermediate results can also be useful. However, application-initiated checkpoint is an application-aware technique and enables two important features: 1)~applications that interact with large data often have a small memory footprint before and after the job, which makes the checkpointing task faster, and makes the CMI's size feasible for migration over the network; and 2)~more control and awareness by the application where it's safe to checkpoint with a minimal side effect.

\bigskip\noindent
\textbf{Question 3.} How to keep a CMI's size small?

We either migrate CMIs over the network or write on a shared disk for NavP. Amazon S3 buckets are often not the fastest I/O option because of their physical location. Therefore, it is important to keep CMIs small in size. The size directly depends on the memory footprint (and associated files if one wants to checkpoint open files as well). One solution is to make the memory footprint small by zeroing out or unmapping unused memory within process address space at the time of checkpoint. However, this would require changes in the application, and the solution would no longer remain transparent.

% Yes, this can work.  - Gene
Another solution is to save the  CMIs incrementally by saving only deltas of each consecutive checkpoint.  This would optimize I/O interaction, but this would require an extra step to replay deltas at restart with the current DMTCP open-source code. We checkpoint after a job finishes and before the application loads another data set into memory, in order to ensure a small size for the CMI. \TOOL{} currently replaces the last CMI with the latest CMI.  In addition, incremental checkpointing is an option for future work.

\bigskip\noindent
\textbf{Question 4}. What if the checkpoint task is interrupted?

\TOOL{} guarantees an atomic checkpointing phase. Thus, \TOOL{} makes sure to not replace previous CMIs if the resources were reclaimed in the middle of an ongoing checkpointing phase.

\bigskip\noindent
\textbf{Question 5.} What are some future optimization to this work?

Currently, \TOOL~writes a CMI first on the shared disk, and then it migrates the saved CMI over network. With a small modification to DMTCP, it is possible to directly stream CMIs over the network, in a manner similar to live migration. 
Also, currently, checkpoint and migration is a two-step process. However, consider a scenario in which the checkpointing task is finished, but the migration has yet to finish. This can lead to an inconsistent state. Therefore, we currently consider {\TOOL.hop} to be complete only once the migration is complete.  

\bigskip\noindent
\textbf{Question 6.} What is the difference between existing adaptive checkpoint and the current work?

The current implementation doesn't take account of previous events, duration, and patterns of EC2 spot instance termination. However, this information can help in choosing a candidate node where \TOOL{} should migrate CMIs to within cloud. One can choose a node that is unlikely to be terminated before a job finishes.

\section{Related Work}
% \note{Twinkle: }

Fault tolerance for long running application in high performance computing (HPC) has been a major area for checkpoint-restart. In recent years, checkpointing is found useful for cost-effective resource utilization within and beyond HPC. Jangjaimon et~al\hbox{.}~\cite{jangjaimon2013effective} discuss effective cost reduction for elastic clouds through adaptive checkpointing. However, the current work does not rely on previous events to checkpoint but focuses on keeping the CMI size small to allow checkpointing at any time.

Yi et~al\hbox{.}~\cite{yi2011monetary} compares several checkpointing schemes including adaptive checkpointing based on hourly, rising edge, basic adaptive and current-price events. However, they assume that the checkpointing cost is known, which is unlikely when the memory footprint varies throughout the workflow. In addition, atomic jobs are not flexible enough to be checkpointed at an arbitrary time. This adds an extra restriction to the existing checkpointing schemes for applications in the cloud.

In order to make checkpoint invocation cost-effective, Jung et~al\hbox{.}~\cite{jung2013vm} predicts termination (out-of-bid) and a price hike for spot instances, based on the history. They use Virtual Machine-based (VM-based) checkpoint and migration to avoid any waiting time at restart. However, migrating a VM instead of an application requires more data transfer over the network, and the migration cost varies based on the network. The current work focuses on reducing the memory footprint to minimize data transfer, while migrating CMIs as data.

To the best of our knowledge, this is the first work to enable the navigational programming model for SDS applications via checkpoint-restart over commercial cloud systems, such as Amazon EC2 spot market.

% Process or job migration techniques also involves checkpoint-restart. 

% There are several papers on using checkpoint-restart mechanism to leverage Amazon EC2 Spot instances. For example,  
% Current SDS practices using atomic tasks as basic ingredients are also worth surveying for comparison purposes. 

\begin{comment} 
\begin{enumerate}
    \item Cost-effective solutions for cloud application to take advantage of Amazon EC2 spot market
    \item Papers related to bringing computation to data
    \item NavP or process-migration related
    \item SDS checkpoint-restart efforts
\end{enumerate}
\end{comment} 

%%% \note{Rough:}
%%% There are several papers on using checkpoint-restart mechanism to leverage Amazon spot instance. For example, Jangjaimon et\hbox{.} al~\cite{jangjaimon2013effective} discusses effective cost reduction for elastic clouds through adaptive checkpointing:
%%% pros: 
%%% cons: uses BLCR that requires root privilege

\section{Conclusion and Future Work}
\label{sec:conclusion}

\begin{comment}
As shown in Fig.~\ref{code:app}, doing FTC in SDS can be achieved with minimal intrusiveness for the application code.
By replacing the $\mathtt{\TOOL.publish()}$ statements in the application with the $\mathtt{\TOOL.hop()}$ statements, the application is automatically transformed into a distributed sequential computing (DSC) program.  Multiple DSCs can then be used to construct ``mobile pipelines'' to achieve parallelism~\cite{pan2004navp}.

This paper demonstrated that application deployment is dynamically done by inserting $\mathtt{\TOOL.hop()}$ statements in the application code, and therefore is done by the scientist programmers ``at their leisure''. This is unlike the current practice in which new code is deployed in the constant integration (CI) process where the SDS experts are immensely involved.
\note{Gene's comment: Pls rewrite, What is meant by the ``something is done by" ... ``at their leisure'"?  Did you mean "continuous integration (CI)" instead of "constant"?  Perhaps change the last part to:  "... (C) process, in which external SDS experts must be intimately involved in the scientific programming."}
\end{comment}

This is an initial attempt in applying Navigational Programming, enabled with process checkpointing by DMTCP, to NASA/JPL science data processing. We have successfully built the NavP Bridging Services (NBS) and demonstrated its application for JPL Satellite Data Co-location processing. 
With our current implementation, we are able to successfully checkpoint, migrate, and resume our app. We also mimicked successfully the continuation of a job on a new instance with intermediate partial results published during the most recent checkpointing. Initial performance studies showed large overhead in our current implementation. We plan to customize and optimize DMTCP, and publish case studies and performance analysis, in a future version of this report.

This report introduced and described several concepts in the NavP paradigm, and provided a rationale
behind why this is superior to the currently prevailing methods. It has also raised many questions.
More case studies are necessary to provide experimental evidence.
With an optimized NBS/DMTCP system in the future, we plan to attack more JPL science data processing problems, and answer the questions that were raised in this report.

\section*{Acknowledgment}
The authors wish to acknowledge useful discussions on this topic with Gene Cooperman.

% \IEEEtriggeratref{6}
\bibliographystyle{IEEEtran}
\bibliography{navp-dmtcp}

\end{document}